# Could the GSI Oscillations be Observed in a Standard Electron Capture Decay Experiment?

Thomas Faestermann[a*], Fritz Bosch[b], Ralf Hertenberger[c], Ludwig Maier[a], Reiner Krücken[a] and Georg Rugel[a]

[a] Physik Department, Technische Universität München, D85748 Garching, Germany
[b] Gesellschaft für Schwerionenforschung, D64291 Darmstadt, Germany
[c] Fakultät für Physik, Ludwig-Maximilians-Universität München, D85748 Garching, Germany

## Abstract

The electron-capture decay of $^{180}$Re has been investigated to search for oscillations in the decay probability as reported from a recent measurement at GSI, Darmstadt. The production period was kept short compared to the reported oscillation period. No such oscillation was observed, indicating that the reported oscillations would not have been observable in a conventional experiment with radioactive atoms in a solid environment but must have to do with the unique conditions in the GSI experiment where hydrogen-like ions are moving independently in a storage ring and decaying directly by a true two-body decay to a long-lived (ground-) state. Our finding could restrict possible theoretical interpretations of the oscillations.

[*] Corresponding author:

postal adress: Thomas Faestermann, Physik Department E12, James-Franck-Str., D85748 Garching, Germany

E-mail adress: thomas.faestermann@ph.tum.de

tel: +49-89-28912438   fax:+49-89-28912297

## 1. Introduction

Recently, measurements have been reported where the decay probability in electron capture (EC) decays of $^{140}$Pr and $^{142}$Pm is not constant with time but has an oscillatory behaviour [1], with a frequency in the lab system for both systems of $\nu = 0.142$ s$^{-1}$. The radioactive hydrogen-like ions were coasting in the experimental storage ring of GSI, Darmstadt, with a kinetic energy of 400·A MeV, corresponding to a Lorentz factor $\gamma = 1.43$. It has been suggested that these oscillations could be caused by the interference between different neutrino mass eigenstates in the two-body decay channel. Also quantitative theoretical calculations for such oscillations have been published [2], immediately followed by a number of theoretical papers, partly favouring this interpretation [3-6], partly rejecting the possibility that neutrino properties could be the cause of such oscillations [7-9]. If the hypothesis of neutrino oscillations however is true, the oscillation frequency observed [1] for both systems in the center of mass system, $\nu = 0.203$s$^{-1}$, should be proportional to $\Delta(m^2)/M$, where $\Delta(m^2)$ is the difference in the squared masses of two neutrino mass eigenstates and M is the mass of the recoiling daughter nucleus. There are of course special conditions under which these experiments have been performed. The ions undergoing electron capture decay had just a single electron and were coasting in a storage ring. They were produced in a time span short compared to their lifetime and to the oscillation period of 7 seconds.

In this letter we address the question why such periodic modulations have not earlier been observed in conventional measurements. Most beta decay half-lives have been measured in beam-off periods after an irradiation period comparable to the lifetime, e.g. after isotope separation, gas jets or other transport devices. Therefore the possibility arises, that such oscillations could be present also in the decay of atoms or low charge state ions under normal conditions but could have been wiped out by production times long compared to the oscillation period. We are not aware of measurements where explicitly short irradiation times had been used and therefore devised an experiment to test whether such oscillations are observable under normal conditions.

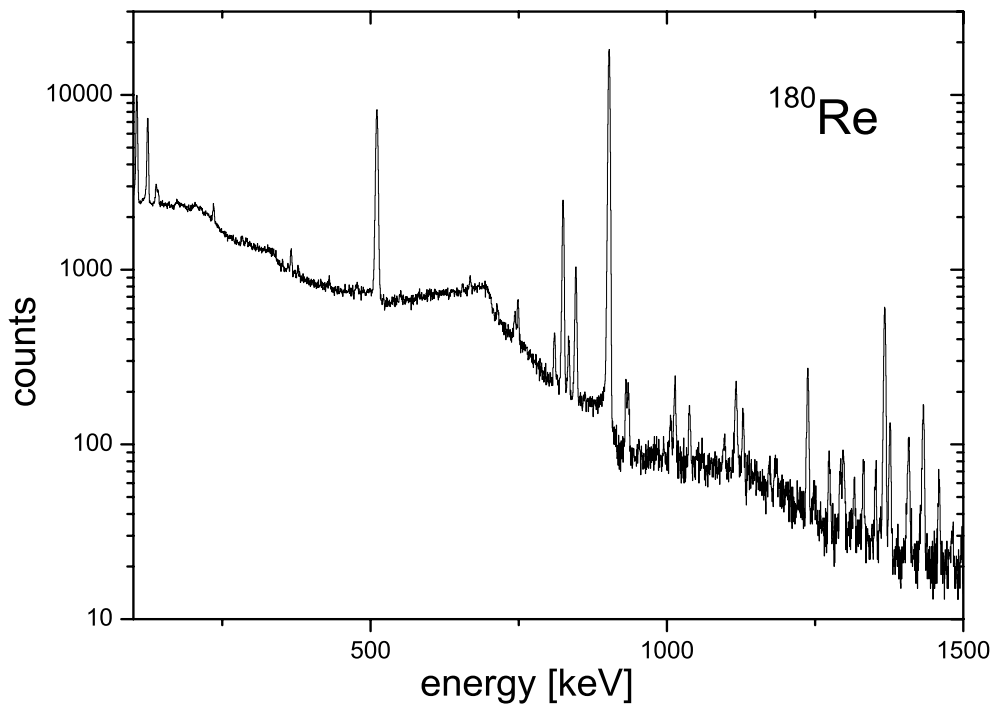

Fig. 1: Electron-capture delayed gamma-spectrum after production of the $^{180}$Re activity with the $^{181}$Ta($^3$He,4n) reaction. The 903 keV gamma-ray is used for the timing of the decay.

## 2. Experimental details

The radionuclide to be studied should fulfil several conditions: It should have a large production cross section in order to produce a large activity already in short irradiations. Its half life should be long compared to the typical oscillation periods of the order of 10s. It should have a large fraction of EC decay, what points in the direction of not too high a Q-value and large Z. Since the neutrino is the only radiation emitted in the decay we have to look for succeeding radiation unlike in the storage ring experiment [1], where the decay has been detected by the change in mass from mother to daughter ion. EC decay is normally followed by x-ray or Auger-electron emission, in case the decaying ion is not highly charged. For experimental reasons we however preferred a succeeding gamma radiation. Therefore we selected $^{180}$Re [10] with a half life of 2.44(6) min as the radioactive isotope for our study. Its $1^-$ ground state decays with 70% probability via an allowed EC decay to a $2^-$ state in $^{180}$W, which then is de-excited to 99% by an E1 transition of 903 keV to the $2^+$ state. This gamma-ray was chosen as the signal indicating the decay. The $2^-$ state in $^{180}$W is in addition fed by 7.2% through $\beta^+$ decay and by 14% from higher levels, mainly through EC decay. In total a fraction of 90% of the 903 keV $\gamma$-transitions is due to EC and thus a two-body decay.

$^{180}$Re has been produced with the $^{181}$Ta($^3$He,4n) reaction. A 50 mg/cm$^2$ tantalum foil was irradiated with a 33 MeV $^3$He beam from the Munich MLL tandem accelerator. The energy

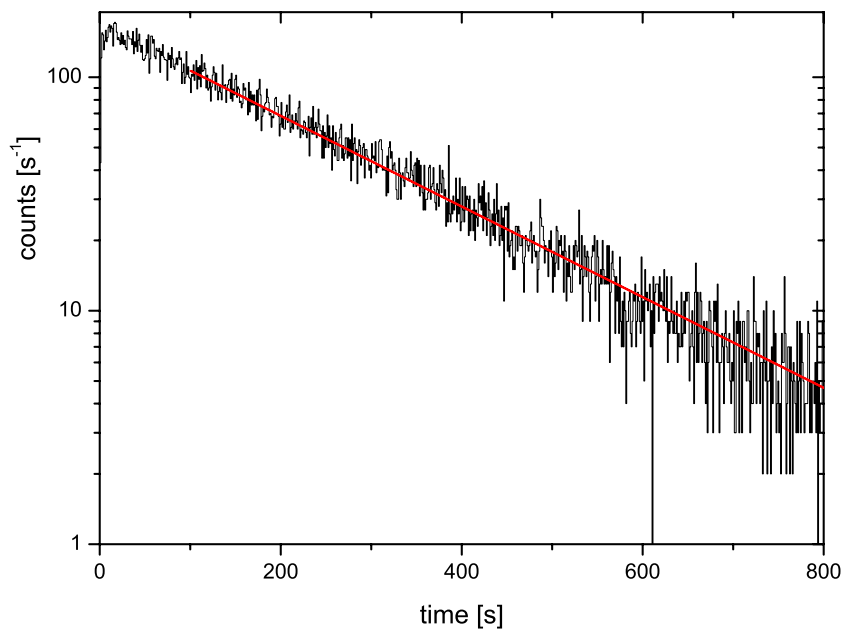

Fig. 2: Time spectrum for $^{180}$Re decays obtained after a single irradiation by setting a narrow gate on the 903 keV gamma line. The time axis is binned with 1 s/bin. The result of a fit is also shown.

loss in the target amounts to 3.5 MeV. $^3$He$^{2+}$ beam currents from the ECR source with charge exchange [11] were between 100 and 500 nA on target. The average production cross section in the target is about 0.2 b [12]. Several irradiations have been performed by removing and inserting a Faraday cup in front of the target after 0.5 s to 1.0 s. A Ge detector at a distance of 20 cm from the target was used to measure the gamma energies and time. A gamma spectrum after a longer irradiation of 23 s duration with a beam current of 60 nA is shown in fig. 1. The 903 keV gamma ray is the strongest line. The peak to background ratio is about 100:1.

Decay time spectra are generated by setting a narrow gate on the 903 keV line. During the irradiation the preamplifier of the Ge detector is overloaded and therefore no events are recorded. This gives the possibility to precisely determine the irradiation time. Still for a short time after irradiation (about 10s) the dead time is high and thus the time spectrum not exponential. This of course does not hamper the search for an oscillation. A part of such a time spectrum after one 0.9s long irradiation is shown in fig.2. The time is measured relative to the middle of the irradiation period. The spectrum from 100-800 s is fitted by an exponential decay with a half life of 2.47(3) min, consistent with the previously known value.

## 3. Results

Five such irradiation cycles have been done. The time between irradiations was at least 15 half lives to let the $^{180}$Re activity die down to a negligible level. The time spectra have been analyzed first by a Fourier transformation to search for possible oscillations. A Fourier transform of one of the spectra (that of fig.2) is shown in fig.3. No prominent frequency line

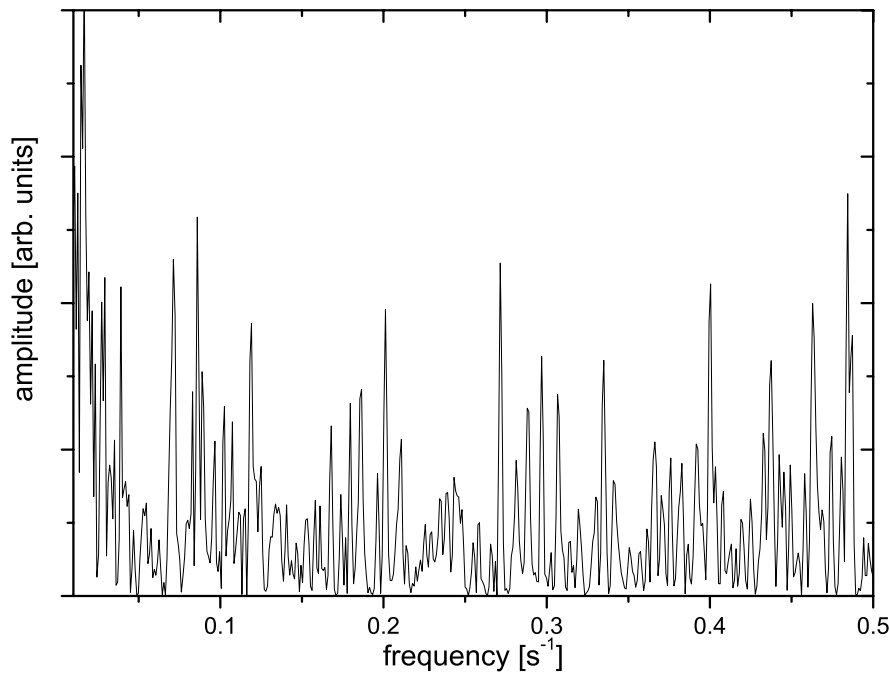

Fig. 3: Fourier transformation of the time spectrum of fig. 2. The GSI oscillation corresponds to a frequency in the A=180 system of $\nu_{180} = 0.158$ s$^{-1}$, supposed that the frequency changes in inverse proportion to the nuclear mass M, as suggested in [1].

was observed in any of the spectra. We then went on to fit the function

$$f(t) = N \exp(-\lambda t) \cdot (1 + a \cdot \cos(2\pi \nu t + \varphi))$$

to the exponential part of the spectra from 10-500 s after irradiation. Including a constant background did not improve the fits. As start values for the frequency ν we took values suggested by the Fourier transform as well as the value from the observation [1], $\nu_{140}$ = 0.203 $s^{-1}$ for mass 140 in the system of the moving ion, scaled by the ratio of recoil masses 140:180, $\nu_{180}$ = 0.158 $s^{-1}$. A fit with just an exponential decay yields $\chi^2$ = 458.7 with 489 degrees of freedom (dof). The data, where the exponential part has been divided out, and two fits with oscillations are shown in Fig.4. The fit with an initial frequency corresponding to the GSI value yields a minimum in $\chi^2$ (456.2 with 486 dof) at a frequency of ν = 0.1582 $s^{-1}$ with an amplitude a=0.013(8). However we do not regard this amplitude as statistically significant, since we are bound to find oscillations at some arbitrary frequencies with amplitudes larger than their error. As largest amplitude we obtain a = 0.028(8) at ν = 0.0715 $s^{-1}$ with $\chi^2$ = 446.2 and 486 dof. We also do not observe oscillations with consistent frequencies in the other runs we did. In addition the amplitude in our measurement for the corresponding frequency is at least a factor of 10 smaller than observed in ref. [1]. Since 90% of the 903 keV γ-transitions originate from EC-decay (see above), we expect a reduction only of that order. The finite length of irradiation would not reduce the amplitude to a great extent. Folding an oscillation of ν = 0.158 s$^{-1}$ with a rectangular function of 1 s length results in a reduction of the amplitude by 4% only.

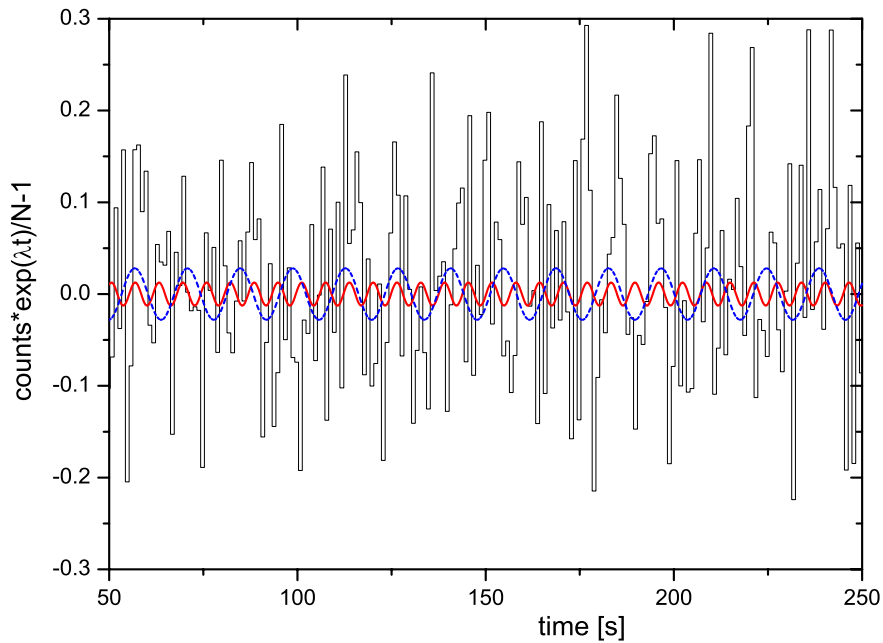

Fig. 4: Part of the time spectrum of fig.2, where the exponential decay has been divided out. The results of fits are also shown: that with a fitted frequency $\nu_{180}$ = 0.1582 $s^{-1}$, that corresponds to the GSI frequency (scaled by the inverse ratio of the masses of the corresponding daughter nuclei), which has a small amplitude (full drawn), and that with the largest amplitude found and a frequency $\nu_{180}$ = 0.0715 $s^{-1}$ (dashed). None of the fits is statistically significant.

Therefore we have to state that in our experiment on EC decay we do not observe the oscillatory behaviour of the ESR experiment at GSI [1]. That implies that previous standard decay experiments could also not see such an effect. This result is, however, not in direct contradiction to the GSI observation [1], since the experiment presented here differs from the GSI experiment in many respects. The state populated by the EC decay in our case is short lived, $T_{1/2}$ = 7.4 ns, and thus the transition energy has a width of the order of $10^{-7}$ eV. The recoiling daughter atom has not a well defined momentum, because it moves in a lattice and can only assume momenta allowed by the phonon spectrum. Therefore the effects causing oscillations in the EC decay to bare ions coasting in a storage ring [1] could be washed out in a standard experiment.

However our non-observation of oscillations in the EC decay probability for a system of atoms in a solid decaying to a short lived excited state might restrict theoretical interpretations of the GSI oscillations.

We want to add that a similar study [13], observing no oscillations in EC decay, has just been made public.

**Acknowledgement**

This investigation was supported by the Maier-Leibnitz-Laboratorium der Münchner Universitäten.